\documentclass[12pt]{iopart}

\usepackage[english]{babel}
\usepackage{amsfonts}
\usepackage{amssymb}
\usepackage{amstext}
\usepackage{epsfig}

\newcommand{\beq}{\begin{equation}}
\newcommand{\eeq}{\end{equation}}
\newcommand{\barr}{\begin{eqnarray}}
\newcommand{\earr}{\end{eqnarray}}
\newcommand{\Ham}{\mathcal H}

\begin{document}
\title{Phase diagram of the extended Bose Hubbard model}

\author{Davide Rossini and Rosario Fazio}

\address{NEST, Scuola Normale Superiore and Istituto Nanoscienze-CNR, I-56126 Pisa, Italy}

\date{\today}

\begin{abstract}

By means of the Density Matrix Renormalization Group technique, we accurately determine 
the zero-temperature phase diagram of the one-dimensional extended Bose Hubbard model
with on-site and nearest-neighbor interactions.
We analyze the scaling of the charge and of the neutral ground-state energy gaps,
as well as of various order parameters. In this way we come to an accurate location
of the boundaries between the superfluid and the insulating phases.
In this last region we are able to distinguish between the conventional Mott insulating 
and density-wave phases, and the Haldane Insulator phase displaying long-range string ordering,
as originally predicted by E.G. Dalla Torre, E. Berg and E. Altman in Phys. Rev. Lett. {\bf 97}, 260401 (2006).

\end{abstract}


% PACSs:
%   05.30.Jp  Boson systems
%   03.75.Hh  Bose-Einstein condensation, static properties
%   64.70.Tg  Quantum phase transitions

\section{Introduction}

In the last decade, ultracold atomic gases loaded in optical lattices successfully established 
as an excellent setup to probe the equilibrium, as well as the out-of-equilibrium physics of strongly 
correlated quantum systems. The great advantages of these experimental setups are essentially related to two aspects.
On one hand, they have an extremely high degree of flexibility:
besides the ability to address different geometries, and to deal with Bosonic and with Fermionic species,
they admit the possibility to manipulate the underlying Hamiltonian system parameters to a large extent.
Moreover, the remarkably high degree of isolation from any environmental source of decoherence 
opened up entirely new scenarios in the observation of genuinely many-body quantum phenomena~\cite{Bloch2008}.

The paradigm model to describe cold Bosonic atoms trapped in an optical lattice 
is obtained by combining the kinetic energy in the lowest band with the on-site repulsion arising 
for sufficiently deep lattices. This leads to the celebrated Bose-Hubbard model (BHM)~\cite{Jaksch1998}.
The rich physics of the BHM stems from the competition between the kinetic energy $J$, 
which is gained by delocalizing particles across the lattice in an extended Bloch state, 
and the repulsive on-site interaction $U$, which disfavors having more than one particle per site.
When the kinetic energy term dominates, the system is in a coherent Superfluid (SF) phase;
on the other hand, repulsive interactions tend to favour a Mott insulating (MI) phase~\cite{Fisher1989}.

The recent advances in manipulating magnetic atoms and molecules with large dipole momentum
make it possible to achieve longer-range interactions, which can be accurately tuned as well,
thus permitting to probe the interplay between strong correlations 
and charge-ordering effects~\cite{Lahaye2009}. 
Dipolar bosons confined in optical lattices are typically described by an extended version 
of the BHM, the so called Extended Bose Hubbard Model (EBHM), which also includes 
a two-body non local repulsive term typically decaying as $r^{-3}$ with the distance $r$.

Interestingly the presence of long range interactions noticeably enriches the phase diagram of the BHM,
for example leading to a stabilization of a peculiar insulating phase,
named the Bosonic Haldane insulator (HI).
This gapped phase presents some analogies with the well known Haldane phase in integer spin chains~\cite{Haldane1983};
it does not break the translational symmetry of the lattice and is characterized 
by an underlying hidden order, that is a non-trivial ordering of the fluctuations 
which appear in alternating order separated by strings of equally populated sites 
of arbitrary length~\cite{DallaTorre2006, Berg2008, Amico2008}.
More recently a mean field analysis suggested the existence of various supersolid 
phases and a series of non-trivial density-wave states with an increased ground-state 
degeneracy, which can be stabilized in the strong coupling regime, but with a relatively 
weak on-site interaction~\cite{Wikberg2012}.
It has been also shown that the presence of disorder in the on-site potential~\cite{Deng2012}, 
as well as occupation-dependent hopping and pair tunneling terms arising 
from dipolar interactions~\cite{Sowinski2012}, can drive important modifications 
in the phase diagram leading to the stabilization of a plethora of novel quantum phases, 
such as structured insulating states, Wigner crystals, or pair-supersolids.

The aim of this paper is to work out the zero-temperature phase diagram 
of the pure one-dimensional (1D) EBHM, by means 
of the numerical Density Matrix Renormalization Group (DMRG) technique~\cite{Schollwock2011},
in order to accurately determine the phase diagram first proposed in Refs.~\cite{DallaTorre2006, Berg2008} 
and extend their preliminary results.
This method has been already revealed its great potentialities in the study of ground states
of 1D strongly correlated bosons on a lattice, providing very accurate results for these kinds 
of systems (see e.g. Refs.~\cite{Monien1998, Kuhner2000, Rapsch1999, Rizzi2005, CoupCav2007, CoupCav2008}
and references therein).

The paper is organized as follows.
In Sec.~\ref{Sec:Model} we introduce the model we are going to study, and qualitatively discuss
the various emerging quantum phases.
We then present, in Sec.~\ref{Sec:quantities}, all the relevant quantities we will address:
i) three kinds of local and non local correlators, which will enable us 
to locate all the different phases identifying suitable order parameters; 
ii) the ground state charge and neutral energy gaps,
which will serve to discriminate critical phases as well as insulating QPT points.
A phase diagram of the model is shown in Sec.~\ref{Sec:PD},
followed by a detailed discussion of the finite-size scaling of the various order parameters 
(SubSec.~\ref{Sec:OrderP}) and of the charge and energy gap (SubSec.~\ref{Sec:Gap}),
up to a few hundreds of sites.
Finally, in Sec.~\ref{Sec:Concl} we draw our conclusions.

\section{The model}  \label{Sec:Model}

In the following we will focus on the 1D extended Bose Hubbard model (EBHM), 
with on-site and nearest-neighbor repulsive interactions.
The sharp cutoff of the $r^{-3}$ interaction range, although presenting quantitative differences
with the standard long-range EHBM, does not qualitatively alter its physics:
\beq
   \Ham = -J \sum_j \left( b_j^\dagger b_{j+1} + H.c. \right) + \frac{U}{2} \sum_j n_j (n_j-1) + V \sum_j n_j n_{j+1} \,;
   \label{eq:EBHM}
\eeq
here $b^\dagger_j, \, b_j$ are creation and annihilation operators of bosons on site $j$,
$J$ denotes the hopping strength, while $U$ and $V$ are the on-site and the nearest-neighbor interaction strengths.
Hereafter we set the energy scale by taking $J=1$, and work in units of $\hbar = 1$.

We will concentrate on the model in Eq.~(\ref{eq:EBHM}) at zero temperature and at integer filling $\bar{n}$, 
which is known to exhibit a quite rich phase diagram with various quantum phases, ranging from the superfluid (SF) 
for low $U,V$ interaction strengths, to insulating phases in the opposite regime.
Depending on the relative strength of $U$ and $V$, a Mott insulator (MI) or a density-wave state (DW) 
can form, the first one establishing with large values of $U$, while the second one with large $V$.
In between the two conventional insulating phases, a peculiar gapped phase emerges: the Haldane insulator (HI).

The phase diagram and the phase transitions of such kind of model, describing bosons loaded 
in optical lattices with short range interactions, have been addressed since long time~\cite{Fisher1989}.
Emphasis was put on the experimentally accessible SF-MI transition~\cite{Greiner2002}.
In the 1D case and with a constant integer filling, the transition is of the Berezinskii-Kosterlitz-Thouless 
type and was extensively analyzed with DMRG in Refs.~\cite{Monien1998, Kuhner2000}.
The same authors also demonstrated that, in presence of nearest-neighbor interactions, there is no normal 
or supersolid phase, but a direct phase transition from the DW to the SF phase occurs.
Only a few years ago, by means of a mean field theory describing low-energy excitations in the continuum limit, 
in Refs.~\cite{DallaTorre2006, Berg2008} it was realized that extended interactions can stabilize 
a further HI phase, possessing hidden order that is revealed by non-local string correlations. 
In those papers the DMRG was employed to extract some preliminary quantitative results 
and draw the resulting phase diagram in 1D.
Further DMRG calculations have been also used to address closely related effective spin-1 chains, 
including finite-size scaling studies~\cite{Ercolessi2003, Ueda2008},
the effects of trapping potentials~\cite{Dalmonte2011}, and to study the entanglement
spectrum in order to probe the degeneracies in the HI phase~\cite{Santos2011}.

Here we numerically probe the EBHM over all its parameters space, corroborating
the investigations previously done in the literature and providing
an accurate description of its complete phase diagram at integer filling.
In order to locate the different quantum phases of Eq.~(\ref{eq:EBHM}) 
we resort to the DMRG algorithm with open boundary conditions~\footnote{
The choice of open boundary conditions is popularly used in the DMRG community,
and it is dictated by the fact that it has been proven that they perform 
more efficiently than periodic boundaries~\cite{Schollwock2011}.
We will take special precautions for border effects emerging in our simulated finite-size systems.}.
Our code is based on a variational ansatz using Matrix Product States (MPS),
in the restricted subspace of integer filling~\cite{Schollwock2011}.

\section{Correlation functions, order parameters and energy gaps}  \label{Sec:quantities}

The nature of the different quantum phases emerging in the EBHM, as well as all the 
Quantum Phase Transition (QPT) points determining the various phase boundaries,
are revealed by the behavior of certain two-point correlation functions. 
The following three correlators are able to univocally distinguish between 
all the emerging quantum phases:
\barr
\label{eq:SF}      C_{\rm SF}(r)    & = & \langle b^\dagger_j b_{j+r} \rangle \,, \\
\label{eq:DW}      C_{\rm DW}(r)    & = & (-1)^r \langle \delta n_j \delta n_{j+r} \rangle \,, \\
\label{eq:string}  C_{\rm string}(r) & = & \langle \delta n_j e^{i \pi \sum_{j \leq k < j+r} \delta n_k} \delta n_{j+r} \rangle \,,
\earr
where $\delta n_j = n_j - \bar{n}$ denotes the boson number fluctuations from the average filling
(hereafter we will always consider the case $\bar{n} = 1$).

The long range off-diagonal order, typical of superfluid states, manifests itself 
into a power-law decay to zero of $C_{\rm SF}(r)$;
on the other hand, in the insulating phases the absence of such ordering
is characterized by an exponential suppression of $C_{\rm SF}(r)$ with $r$.
The large-$r$ limit of the second correlator, $C_{\rm DW}(r)$, identifies
a staggered diagonal order which naturally emerges in the DW phase:
at $\bar{n} = 1$, large values of $V$ in Eq.~(\ref{eq:EBHM}) tend to stabilize such ordering,
by inducing a pattern $\ldots -2-0-2-0- \ldots$ for the occupation number.
Such phases are then characterized by a finite DW order parameter 
${\cal O}_{\rm DW} \equiv \lim_{r \to \infty} C_{\rm DW}(r)$. 
Finally, the two-point correlator $C_{\rm string}(r)$ contains a non-local string correlation function
accounting for the number of bosons that are present in between the two points.
For large values of $r$, this has been shown to define an order parameter for the
HI, namely ${\cal O}_{\rm string}\equiv \lim_{r \to \infty} C_{\rm string}(r)$, 
which distinguishes from the usual MI though such highly non-local ordering.
On the contrary, in the incoherent MI phase all the correlation functions 
in Eqs.~(\ref{eq:SF})-(\ref{eq:DW})-(\ref{eq:string}) vanish exponentially.

In addition to the previously defined correlation functions, we also analyze the energy gap.
The nature of the different quantum phases, as well as the QPT points are indeed sensitive to it.
In particular, we will focus on the neutral ($\Delta E_n$) and on the charge ($\Delta E_c$)
gaps, respectively defined as:
\barr
\label{eq:GapN}  \Delta E_n & = & E^{(1)}_L - E^{(0)}_L  \\
\label{eq:GapC}  \Delta E_c & = & E^{(0)}_{L+1} + E^{(0)}_{L-1} - 2 E^{(0)}_{L} \, ,
\earr
where $E^{(k)}_m$ indicates the energy of the $k$-th excited state ($k=0$ labelling the ground state,
$k=1$ the first excited state, and so on), for a system with a given number of $m$ bosons 
in $L$ lattice sites~\footnote{
The case $m=L$ corresponds exactly to the integer filling sector $\bar{n} = 1$,
in which the neutral gap is evaluated.
The charge gap in Eq.~(\ref{eq:GapC}) expresses the difference 
in the energy costs $\Delta E^{(0)}_\pm$ to add and to subtract a single particle 
from the ground state at unitary filling, i.e. 
$\Delta E_c = \Delta E^{(0)}_+ - \Delta E^{(0)}_- = (E^{(0)}_{L+1} - E^{(0)}_{L}) - (E^{(0)}_{L-1} - E^{(0)}_{L})$.}.
In practice we run our DMRG simulations in the canonical ensemble.
The charge gap is evaluated by running three instances of DMRG, each of them separately targeting
the ground state of $n = \{ L, L \pm 1 \}$ particle number sectors~\cite{Monien1998, Kuhner2000}.
The neutral gap is extracted by running only one instance of DMRG, in which
both the ground state and the first excited state are 
targeted~\footnote{
In general, fixing the system size and the bond-link dimension,
the evaluation of $\Delta E_c$ is much faster and more accurate than that of $\Delta E_n$,
since only the ground state has to be addressed.}.

Unfortunately, the charge gap $\Delta E_c$ only detects particle and hole excitations,
and it is able to locate the superfluid-to-insulator transition, 
as well as the MI-HI QPT point. 
On the other hand, the HI phase can display another type of excitation 
which in energy goes below the particle-hole excitation,
and is revealed only by the neutral gap $\Delta E_n$.
It turns out that the presence of such low-energy neutral modes becomes crucial
for the identification of the HI-DW transition, as we will explicitly show later.

In the next Section we will discuss in details our results.
We first show the phase diagram of the EBHM, distinguishing between the various
quantum phases of the model. Then we will present our data on the order parameters
and the ground-state energy gaps.

\section{Phase diagram}   \label{Sec:PD}

\begin{figure}[t]
  \begin{center}
    \includegraphics[width=13cm]{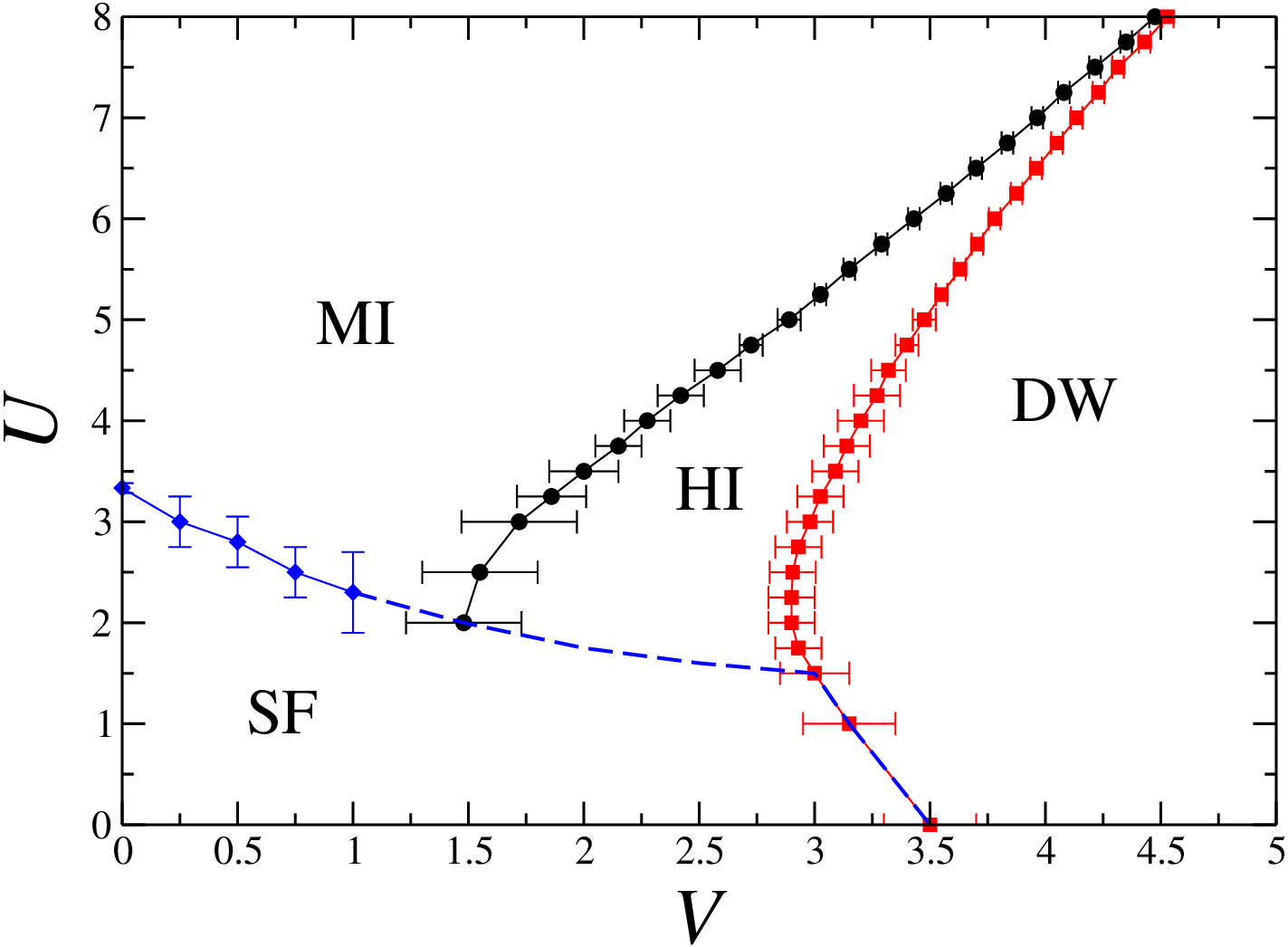}
    \caption{Ground-state phase diagram of the 1D EBHM in the ($U,V$) plane,
      with a maximum admittable number of 3 bosons per site.
      The boundaries between the different phases have been calculated by analyzing 
      the string order parameter (MI-HI --- black circles), 
      the density wave order parameter (HI-DW --- red squares)
      and the charge energy gaps of the system (SF-MI --- blue diamonds).
      The dashed blue line is an interpolation of the continuous blue curve
      and the red points at large $V$ values.}
    \label{fig:PhaseDiag}
  \end{center}
\end{figure}

We start the discussion of our results by presenting the phase diagram of Eq.~(\ref{eq:EBHM}) in the ($U,V$) plane. 
Our DMRG simulations were performed for systems with open boundary conditions up to $L=400$ sites, 
keeping at most $m=500$ states, and working in a canonical ensemble with a fixed number of bosons.
Due to computational reasons, for a filling $\bar{n} = 1$ we admitted a maximum of $n_{\rm max} = 3$ 
bosons per site. 
%For higher cutoffs the emerging phase diagram would not 
%qualitatively change, while this would require much more demanding computational efforts
We checked that the location of the phase boundaries is not qualitatively affected 
by the values of the various cutoffs~\footnote{
We also performed simulations fixing $n_{\rm max} = 2$, thus mapping the problem into a spin-1
Heisenberg Hamiltonian with a single-site uniaxial anisotropy, and explicitly found 
a phase diagram with the same qualitative features as those depicted Fig.~\ref{fig:PhaseDiag}. 
We however noticed important quantitative changes,
with the HI phase being shifted at lower values of $V$, and the DW phase enlarged.}.

%%%%%%%%%%%%%%%%%%%%%%%%%%%%%
%%%%%%%%%%%%%%%%%%%%%%%%%%%%%

Figure~\ref{fig:PhaseDiag} summarizes the results of our investigation.
As we stated before, model~(\ref{eq:EBHM}) has four different quantum phases.
For large $U$ values, the system is always insulating.
In order to distinguish between MI, HI and DW phases, we carefully
analyzed the string and the density-wave order parameter, respectively ${\cal O}_{\rm string}$
and ${\cal O}_{\rm DW}$.
An accurate finite-size scaling of the ${\cal O}_{\rm string}$ data allowed us to
draw the black line (data are shown as circles) in Fig.~\ref{fig:PhaseDiag}.
A similar analysis was also performed on ${\cal O}_{\rm DW}$, producing the red curve 
in the figure (red squares).
For further details in the specific situation with $U=5$ we refer to Sec.~\ref{Sec:OrderP}.
We performed a cross-check of such obtained phase boundaries by looking
at the charge and the neutral energy gaps (see Sec.~\ref{Sec:Gap}),
obtaining minimal values for the gaps at the QPT points.

For small values of $U$ and not too large $V$, the system enters a critical SF phase,
characterized by long-range off diagonal order.
We located the SF by looking at the charge energy gap (blue diamonds).
While for small $V$ values the determination of the SF-insulator boundary is
quite accurate at relatively small sizes $L \approx 300$ (where, by increasing $U$,
at some critical point the charge energy gap undergoes a sudden increase), 
this becomes harder for larger $V$.
The dashed line in the figure has been plotted as a guide to the eye, and is an interpolation
between the blue data in the small-$V$ region and the red DW boundary at large values of $V$.
As a matter of fact, despite the relatively low numerical accuracy that we were able
to achieve in that region, our data are indeed compatible with the presence of a SF phase 
below the HI, for small U values and $1.5 \lesssim V \lesssim 3$, in agreement with 
the findings of Ref.~\cite{Deng2012}.

\subsection{Scaling of the order parameters}  \label{Sec:OrderP}

In order to determine the HI-DW and the MI-HI phase boundaries, we respectively looked at 
the correlation functions in Eqs.~(\ref{eq:DW})-(\ref{eq:string}).
In particular we computed the string and the density-wave order parameters by
extracting the large-$r$ limit of such correlators. To minimize border effects due to
open boundaries, we computed the average expectation values of all the two-point correlators
between spins in the middle part of the chain, discarding the outer $L/4$ sites on both sides.
We also studied the dependence on the system size by performing a finite size
scaling up to $L=400$ system sites.

\begin{figure}
  \begin{center}
    \includegraphics[width=13cm]{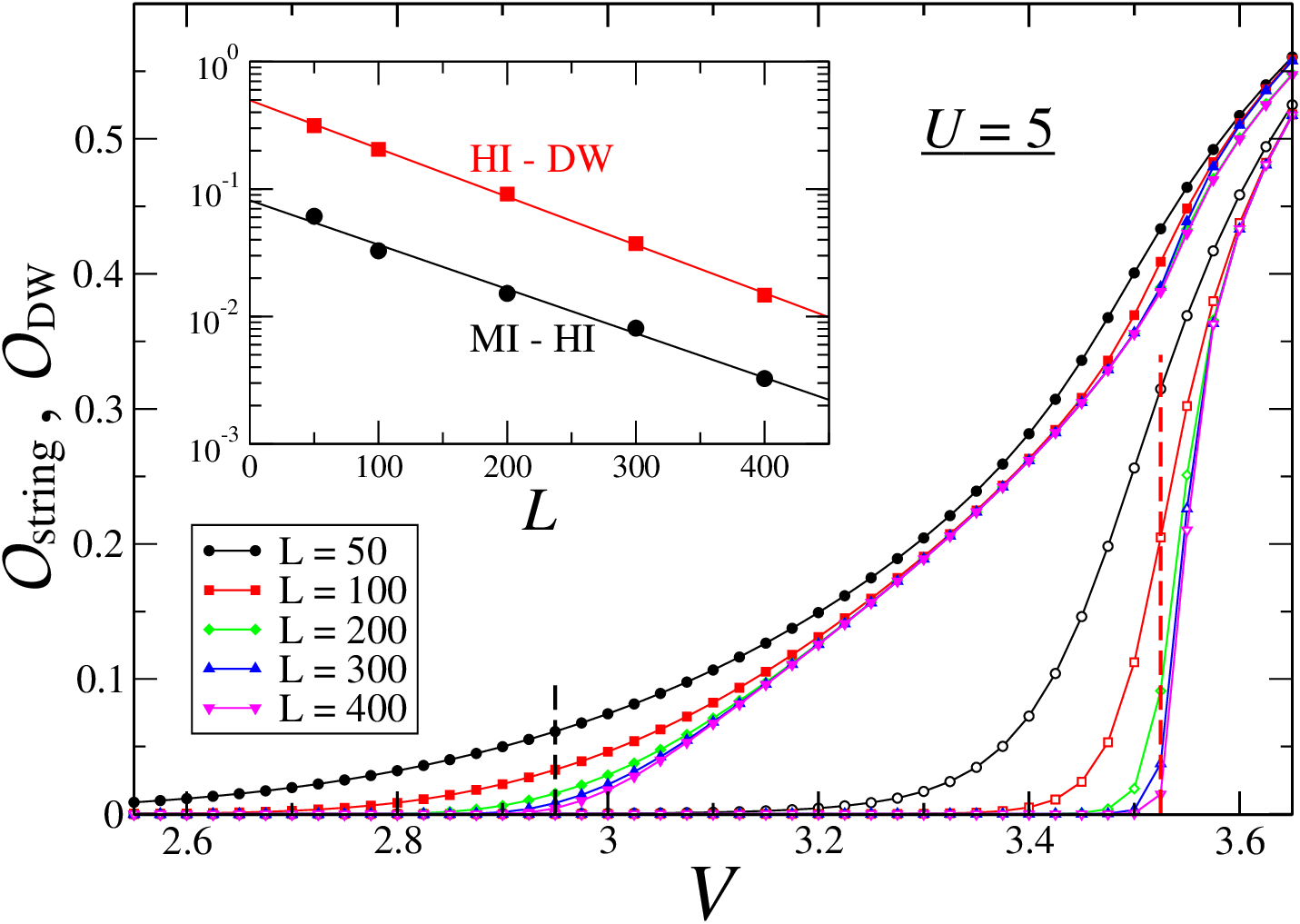}
    \caption{String and density-wave order parameters (${\cal O}_{\rm string}$, ${\cal O}_{\rm DW}$),
      as a function of $V$, for a fixed $U=5$ and for different lengths $L$. 
      The system exhibits a transition from the MI to the HI phase
      at $V \approx 2.95 \pm 0.05$, where the string order parameter starts being finite
      (filled symbols). At larger values of $V$ a further transition from the HI to the DW phase
      is signaled at $V \approx 3.525 \pm 0.05$ (empty symbols, displaying the density-wave order parameter).
      In the inset we plot the values of the two order parameters in proximity of the phase transitions
      (at $V_{\rm MI-HI}=2.95$ and at $V_{\rm HI-DW}=3.525$), showing that 
      they exponentially go to zero with the system size $L$. 
      The straight lines show the best fits of numerical data (symbols): 
      ${\cal O}_{\rm string} = 0.0815 \times e^{-0.008 \, L}$ and ${\cal O}_{\rm DW} = 0.4968 \times e^{-0.0087 \, L}$.}
    \label{fig:OrdPar}
  \end{center}
\end{figure}

An example of such analysis is shown in Fig.~\ref{fig:OrdPar}, where we fixed the on-site
interaction strength $U=5$, and varied $V$.
Both the HI order parameter (filled symbols, on the left part of the main panel),
and the DW order parameter (empty symbols, on the right of the main panel)
are plotted as a function of $V$ and for different sizes ranging from $L=50$ to $L=400$.
The MI-HI phase transition is located at the point where ${\cal O}_{\rm string}$ becomes finite,
which is estimated to be at $V_{\rm MI-HI} \approx 2.95 \pm 0.05$ (dashed black vertical line).
On the other hand, the HI-DW phase transition occurs where a finite ${\cal O}_{\rm DW}$ appears,
that is at $V_{\rm HI-DW} \approx 3.525 \pm 0.05$ (dashed red vertical line).

%%%%%%%%%%%%%%%%%%%%%%%%%%%%%
%%%%%%%%%%%%%%%%%%%%%%%%%%%%%

Finite-size effects are clearly visible from the figure.
In the inset we show how the two order parameters approach the zero value very close to
the QPT. Namely, we plotted ${\cal O}_{\rm string}$ (black circles) and ${\cal O}_{\rm DW}$ (red squares)
respectively at $V_{\rm MI-HI}=2.95$ and $V_{\rm HI-DW}=3.525$.
In both cases we obtained an exponential decay of ${\cal O}$ with the system size $L$.

\subsection{Scaling of the energy gaps}  \label{Sec:Gap}

%%%%%%%%%%%%%%%%%%%%%%%%%%%%%
\begin{figure}[b]
  \begin{center}
    \includegraphics[width=15cm]{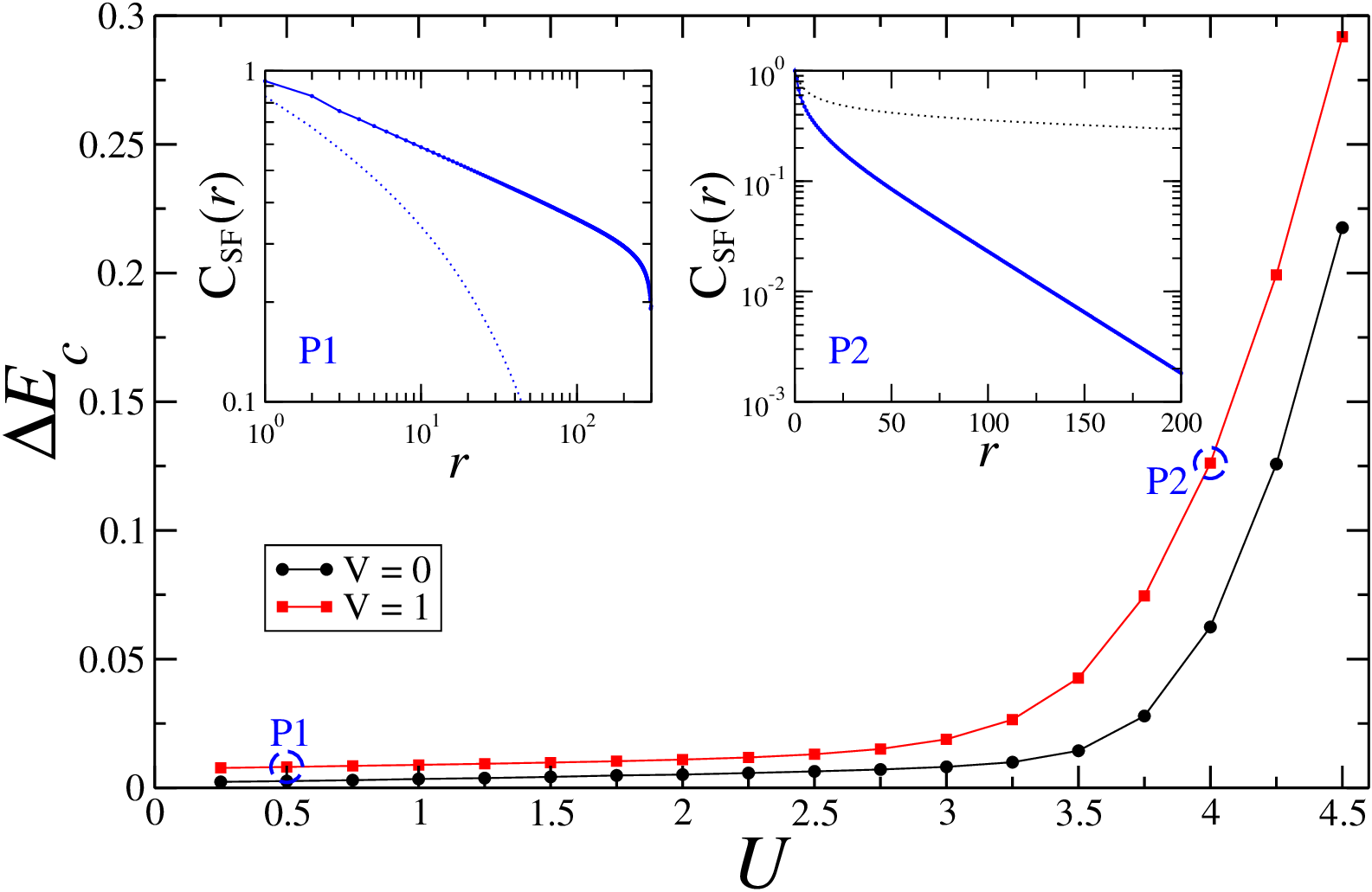}
    \caption{Ground-state charge energy gap $\Delta E_c$ as a function of $U$, for two different values of $V$.
      The two insets display the correlation function $C_{\rm SF}(r)$ as a function of $r$,
      deep in the superfluid phase (left inset -- P1 at $V=1$, $U=0.5$) and deep in the
      Mott insulating phase (right inset -- P2 at $V=1$, $U=4$).
      The dotted lines show the complementary cases (P2 in the left inset and P1 in the right inset),
      and are plotted in order to better highlight the different power-law vs. exponential behavior.
      Data are for a system of $L=400$ sites. The outer $100$ sites have been discarded such to
      minimize boundary effects.}
    \label{fig:Gaps_SF}
  \end{center}
\end{figure}
%%%%%%%%%%%%%%%%%%%%%%%%%%%%%

Further analysis of the QPT points has been performed by looking at the neutral and at the charge
energy gaps, $\Delta E_c$ and $\Delta E_n$ [see Eqs.~(\ref{eq:GapN}),(\ref{eq:GapC})].
In addition to corroborating the results obtained from the order parameters scaling analysis
as we will show in Fig.~\ref{fig:ChargeGap} and Fig.~\ref{fig:NeutralGap},
the closure of the energy gaps discriminates between the SF and the insulating phases.
This is a method to locate the SF boundaries, which reveals more accurate than directly
looking at the off diagonal ordering.
As a representative example, in Fig.~\ref{fig:Gaps_SF} we show the behavior of the
charge energy gap in proximity of the SF-MI transition of model~(\ref{eq:EBHM}), 
as long as $U$ is increased upon a critical value (see blue diamonds in Fig.~\ref{fig:PhaseDiag}).
Namely, we plotted $\Delta E_c$ as a function of $U$, for a fixed value of $V$ and
for a system of $L=400$ sites.
The SF-MI QPT is revealed by a sudden growth of $\Delta E_c$ at a critical value $U^\star$, 
which is estimated to be $U^\star \approx 3.33$ at $V=0$ and $U^\star \approx 2.3$ at $V=1$~\footnote{
The value $U^\star \approx 3.33$ corresponds to the quantum critical point for the SF-MI transition 
in the 1D BH model at filling $\rho = 1$, which has been thoroughly investigated in the past literature 
(see, for example, Ref.~\cite{Kuhner2000}).}.
We remark that, at the sizes that are typically studied, it is easy to discriminate between 
a power-law decay of $C_{\rm SF}(r)$ deep in the SF phase and its exponential decay deep in the MI,
as it is clearly shown in the two insets of Fig.~\ref{fig:Gaps_SF}. 
Nonetheless, this becomes much trickier in proximity of the phase boundaries 
and would require very large sizes to distinguish between the two behaviors~\cite{Kuhner2000}.

We point out that we are always interested in the bulk energy gaps,
even if our DMRG method preferably works with open boundary conditions.
Unfortunately the appearance of edge states in the HI phase can interfere with 
the determination of the bulk gaps. For this reason, we lifted the excitations at the borders 
by forcedly requiring to have zero particles on the leftmost site of the chain, 
and two particles on the rightmost site. This corresponds to apply some strong field at the edges.
Such a bias also breaks the ground state degeneracy of the DW phase, thus
stabilizing the numerical algorithm at large values of $V$.

\subsubsection{Ground-state charge energy gap. \\} 

%%%%%%%%%%%%%%%%%%%%%%%%%%%%%
\begin{figure}
  \begin{center}
    \includegraphics[width=13cm]{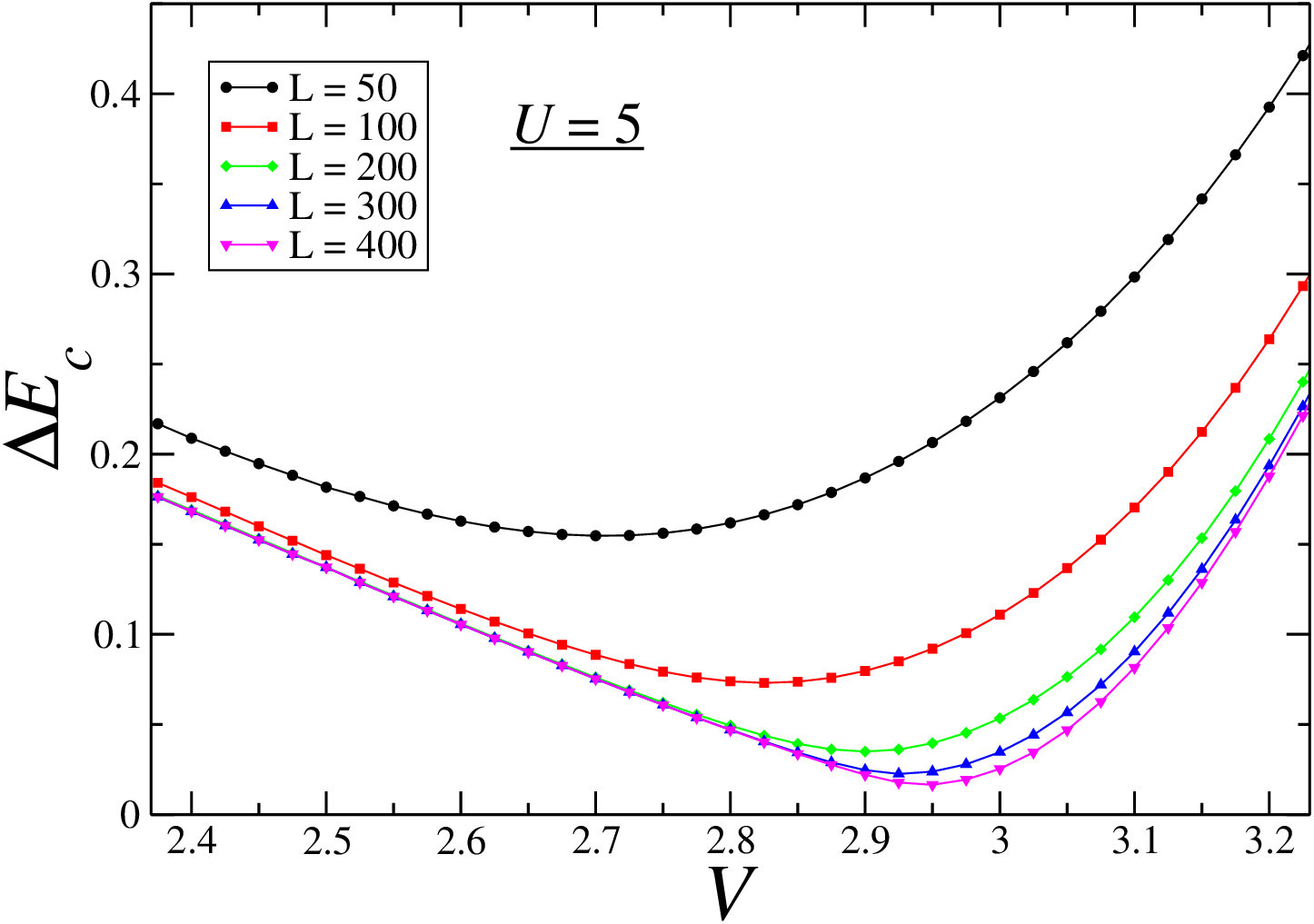}
    \caption{Ground-state charge energy gap $\Delta E_c$ in proximity of the MI-HI phase transition as a function of $V$, 
      for a fixed $U=5$, and for different chain lengths $L$. Data are in accordance to the ones 
      in Fig.~\ref{fig:OrdPar}, showing that the QPT takes place at $V \approx 2.95$.}
    \label{fig:ChargeGap}

    \vspace*{1.5cm}
    \includegraphics[width=13cm]{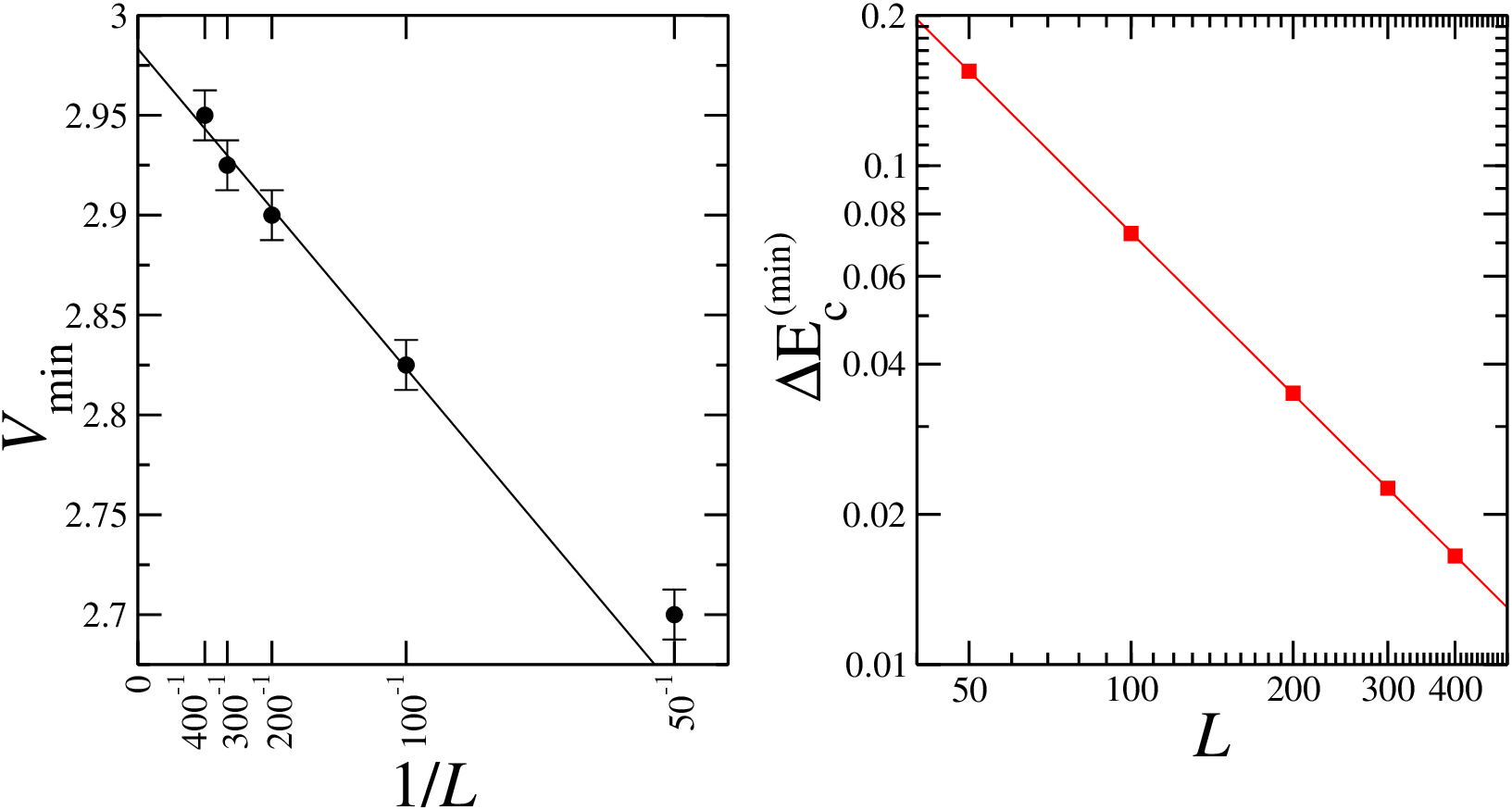}
    \caption{Finite-size scaling of the charge energy gap (data taken from Fig.~\ref{fig:ChargeGap}).
      The two panels display the value of $V_{\rm min}$ at which $\Delta E_c$ reaches its minimum [left panel] 
      and the corresponding $\Delta E_c$ value [right panel: $\Delta E_c^{(\rm min)} \equiv \Delta E_c (V_{\rm min})$].
      In the left panel the continuous black line denotes the linear fit $V_{\rm min} = 2.983 - 16/L$,
      thus extrapolating a value $V_{\rm min}^{L \to \infty} \approx 2.98$ at the thermodynamic limit.
      In the right panel it is shown that the minimum gap closes as a power law at the critical point,
      with a best fitting curve of numerical data given by the continuous red curve: 
      $\Delta E_c^{(\rm min)} = 10.347 \times L^{-1.075}$.}
    \label{fig:Gap_conv}
  \end{center}
\end{figure}
%%%%%%%%%%%%%%%%%%%%%%%%%%%%%

In Fig.~\ref{fig:ChargeGap} we show the charge energy gap as a function of $V$
and for different number of sites $L$, after fixing the on-site interaction strength $U=5$ 
(same value as in Fig.~\ref{fig:OrdPar}).
The non monotonic behavior, with a minimum gap that closes for $L \to \infty$,
indicates the presence of an insulating QPT between a MI phase on the left side, 
and a HI phase on the right side.

A finite-size scaling of the obtained results is performed in Fig.~\ref{fig:Gap_conv}.
In the left panel we plot the corresponding value of $V$ where the gap reaches its minimum,
as a function of the system size.
By linearly fitting our numerical data as a function of $1/L$, we are able to
extrapolate a value at the thermodynamic limit that is equal to $V_{min}^{L \to \infty} \approx 2.98$.
Under numerical accuracies, this is accordance with the value $V \approx 2.95$ of the 
MI-HI QPT point that was found from the analysis of ${\cal O}_{\rm string}$ at $U=5$.
The right panel clearly shows that the minimum charge energy gap $\Delta E_c$, which is reached at the MI-HI transition,
drops to zero as a power law with the size, following the behavior $\Delta E_c \sim 1/L$.
This kind of behavior at the QPT transition can be explained within the conformal field theory approach,
where it has been proven that finite-size corrections to the energies of low-lying gapless excitations 
behave as $1/L$~\cite{Cardy1986, Affleck1986}.

\subsubsection{Ground-state neutral energy gap. \\}

\begin{figure}[t]
  \begin{center}
    \includegraphics[width=13cm]{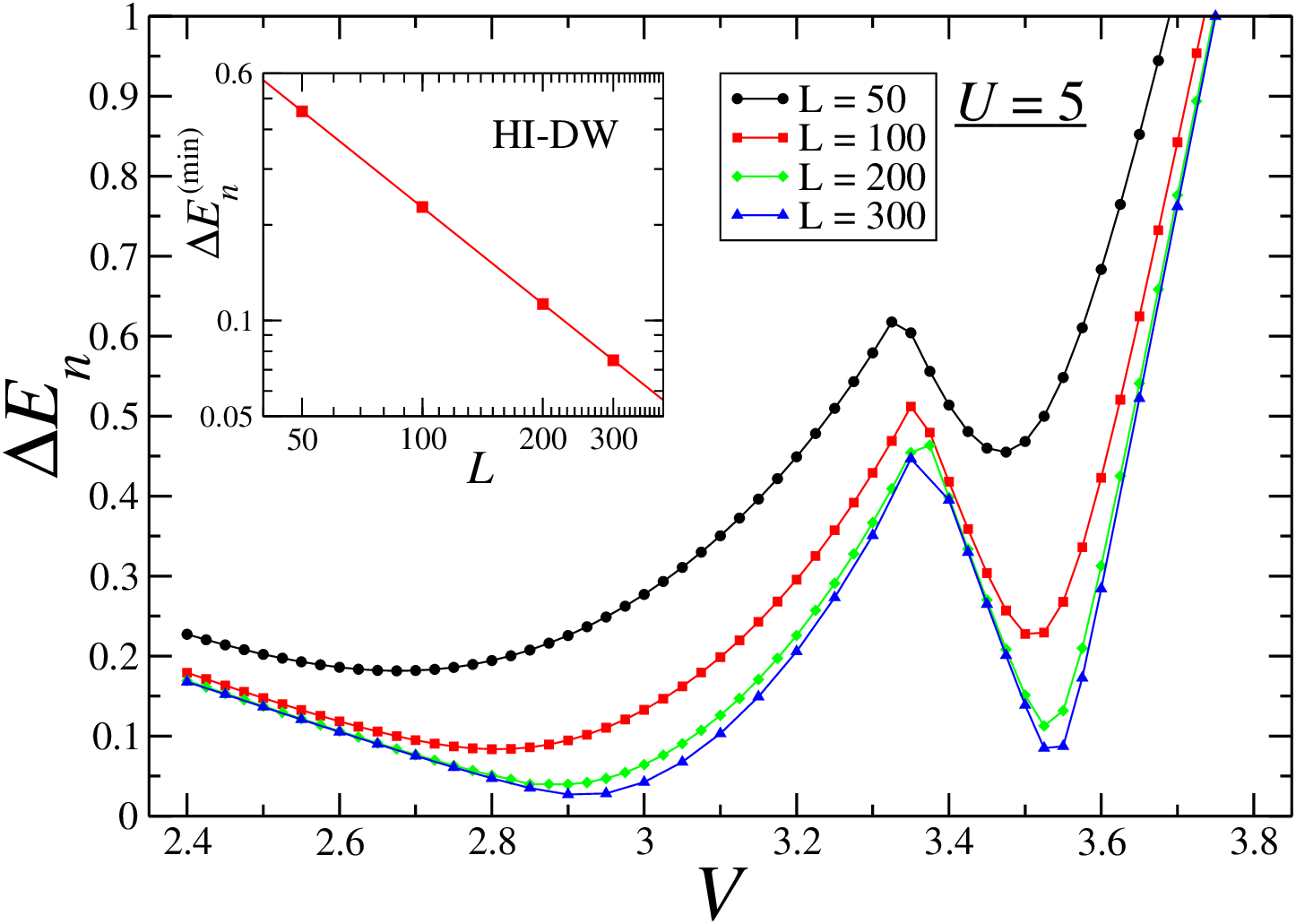}
    \caption{Neutral ground-state energy gap $\Delta E_n$ as a function of $V$, 
      for a fixed $U=5$, and for different lengths $L$. Data are in accordance to the ones 
      in Fig.~\ref{fig:OrdPar}, showing a first QPT at $V \approx 2.95$,
      and also a second QPT at $V \approx 3.5$.
      Inset: finite-size scaling of the charge energy gap in proximity of the second QPT,
      i.e. for the HI-DW transition. Like for the charge gap, the minimum neutral gap is also shown 
      to decay as a power law with $L$, according to the fitting curve: 
      $\Delta E_n^{(\rm min)} = 23.467 \times L^{-1.0074}$ (continuous line).}
    \label{fig:NeutralGap}
  \end{center}
\end{figure}

Let us now concentrate on the neutral gap $\Delta E_n$, which is plotted in Fig.~\ref{fig:NeutralGap}
for the same Hamiltonian parameters of Fig.~\ref{fig:ChargeGap}.
%
%%%%%%%%%%%%%%%%%%%%%%%%%%%%%
%%%%%%%%%%%%%%%%%%%%%%%%%%%%%
%
Besides the QPT between the MI and the HI located at $V \approx 3$,
another critical point at $V \approx 3.5$ emerges, separating the HI and the DW phase
(see also Fig.~\ref{fig:PhaseDiag}).
In the MI phase and in part of the HI phase, the particle and hole excitations 
(detected by $\Delta E_c$) are the lowest-energy excitations.
As it is seen from the figure, this is no longer the case for $V > V^\star \approx 3.3$, 
starting from the middle of the Haldane phase.
The presence of a marked cusp at $V^\star$ indeed indicates a modification in the nature
of the lowest excitation, which becomes of a different kind.
The charge gap is not able to detect such excitation, and the transition to the DW phase,
occurring at $V > V^\star$.
On the contrary, the non monotonic behavior of the neutral gap at $V^*$
and its dependence on the size, clearly indicate it through a gap closure.

In the inset of Fig.~\ref{fig:NeutralGap} we analyze the behavior of 
the minimum neutral gap at the HI-DW transition, showing that, similarly to the minimum charge gap 
at the MI-HI transition, it drops to zero with the size as $1/L$~\cite{Cardy1986, Affleck1986}.

Finally we remark that, for the finite sizes we considered, the neutral gap
does not perfectly coincide with the charge gap at $V < V^\star$.
This slight discrepancy is barely visible from a direct comparison of
Fig.~\ref{fig:ChargeGap} and Fig.~\ref{fig:NeutralGap}, and comes from boundary effects
which are more important for small sizes, and split the difference in the energies 
of the first charge and neutral excitations.
Differences are also induced by the bias added to lift the border excitations.

\section{Conclusions}  \label{Sec:Concl}

In this paper we worked out an accurate phase diagram of the extended Bose Hubbard model in one dimension,
using the DMRG algorithm with open boundary conditions.
We located the superfluid, as well as the Mott insulating, the density wave and the Haldane insulating phases,
by performing an up-to-date finite-size scaling with systems of up to $L=400$ sites.
Our analysis involved the order parameters for the HI and the DW, and 
the study of the ground state charge and neutral gaps.
We point out that it would be tempting to analyze also the bipartite fluctuations of the boson number
in a sub-block of the whole lattice system; this method can be employed within standard DMRG simulations 
and without any additional computational effort. Being able to clearly distinguish
between critical and gapped phases, it could provide more accurate data for the location of
the superfluid region at large values of $V$~\cite{CoupCav2008, Rachel2012}.

Besides giving rigorous predictions for future investigations of strongly correlated quantum phases 
in cold gases of dipolar atoms, our quantitative study is also relevant in the context of
quantized transport through adiabatic pumping~\cite{Thouless1983}.
The model we discussed here indeed supports two distinct insulating phases separated by
a critical point: the Mott and the Haldane insulator.
One can open a gap at such point by breaking the inversion symmetry of the system
(for example, via a correlated tight binding hopping for the bosons).
Therefore, by suitably tuning the various parameters of the Hamiltonian, it is possible 
to adiabatically encircle the MI-HI critical point.
As a consequence, the non trivial topology of the isolated gapless point inside the loop
would induce the transport of one single boson through the chain~\cite{Berg2011}.

\section*{Acknowledgments}

We thank M. Gibertini and V. Giovannetti for fruitful discussions, 
and financial support from EU through projects SOLID and NANOCTM.

%%%%%%%%%%%%%%%%%%%%%%%%%%%%%%%%%%%%%%%%%%%%%%%%%%%%%%%%%%%%%%%%%%%%%%%%%%

\end{document}